\documentclass[10pt,confernce]{IEEEtran}
\makeatletter
\newcommand*\titleheader[1]{\gdef\@titleheader{#1}}
\AtBeginDocument{%
  \let\st@red@title\@title
  \def\@title{%
    \bgroup\normalfont\large\centering\@titleheader\par\egroup
    \vskip1.5em\st@red@title}
}
\makeatother

\usepackage{amsmath,graphicx}

\usepackage{amsmath,amssymb,amsfonts}
\usepackage{algorithmic}
\usepackage{graphicx}
\usepackage{makecell}
\usepackage{subcaption}
\usepackage[table]{xcolor}
\usepackage{multirow}

\title{Brightening the Optical Flow through Posit Arithmetic\vspace{-3mm}}
\titleheader{Preprint - Accepted in ISQED 2021\vspace{-15mm}}
\author{\IEEEauthorblockN{Vinay Saxena\IEEEauthorrefmark{1}, Ankitha Reddy\IEEEauthorrefmark{1}, Jonathan Neudorfer\IEEEauthorrefmark{1}, John Gustafson\IEEEauthorrefmark{3}, \\ Sangeeth Nambiar\IEEEauthorrefmark{1}, Rainer Leupers\IEEEauthorrefmark{2}, Farhad Merchant\IEEEauthorrefmark{2}} \\
		\IEEEauthorblockA{\IEEEauthorrefmark{1}Bosch Research and Technology Centre - India, Bangalore} \\
        \IEEEauthorblockA{\IEEEauthorrefmark{2}Institute for Communication Technologies and Embedded Systems, RWTH Aachen University, Germany} \\
        \IEEEauthorblockA{\IEEEauthorrefmark{3}National University of Singapore, Singapore} \\
        \{vinay.saxena, ankitha.reddy, jonathan.neudorfer, sangeeth.nambiar\}@in.bosch.com, john.gustafson@nus.edu.sg \\
        \{farhad.merchant, leupers\}@ice.rwth-aachen.de \vspace{-12mm}}

\begin{document}
\IEEEtitleabstractindextext{%
}
\IEEEoverridecommandlockouts
\IEEEpubid{\makebox[\columnwidth]{~\copyright2021 IEEE \hfill} \hspace{\columnsep}\makebox[\columnwidth]{ }}
\maketitle
%
%

%
\begin{abstract}
As new technologies are invented, their commercial viability needs to be carefully examined along with their technical merits and demerits. The \emph{posit}\textsuperscript{TM} data format, proposed as a drop-in replacement for IEEE 754\textsuperscript{TM} float format, is one such invention that requires extensive theoretical and experimental study to identify products that can benefit from the advantages of posits for specific market segments. In this paper, we present an extensive empirical study of posit-based arithmetic vis-\`a-vis IEEE 754 compliant arithmetic for the optical flow estimation method called Lucas-Kanade (LuKa). First, we use \emph{SoftPosit} and \emph{SoftFloat} format emulators to perform an empirical error analysis of the LuKa method. Our study shows that the average error in LuKa with SoftPosit is an order of magnitude lower than LuKa with SoftFloat. We then present the integration of the hardware implementation of a posit adder and multiplier in a RISC-V open-source platform. We make several recommendations, along with the analysis of LuKa in the RISC-V context, for future generation platforms incorporating posit arithmetic units.    
\end{abstract}
\section{Introduction}
The \emph{posit} data type is proposed as a drop-in replacement for IEEE 754 compliant floating-point format \cite{gustafson1}. Posit format offers compelling advantages over IEEE 754 compliant float format, such as higher accuracy and wider dynamic range. For arithmetic operations, posits require simpler hardware compared to a fully-compliant implementation of IEEE 754 floats \cite{posit1}\cite{Gustafson2020}. It has been shown experimentally that an $n$-bit floating-point adder/multiplier can be replaced by an $m$-bit posit adder/multiplier where $m<n$, without compromising accuracy and range \cite{posit2}\cite{iccd1}. This is due to greater information-per-bit in the posit data type compared to its IEEE-compliant counterpart. Several researchers around the world are working on the efficient realization of posit arithmetic units; studies of posit arithmetic for different application domains have been published \cite{expand1}\cite{posit3}. The \emph{SoftPosit} emulation library supports float-like arithmetic operations with different posit configurations and is closely patterned after the \emph{SoftFloat} library from Berkeley. We believe the time has arrived to apply SoftPosit and SoftFloat to analyze the merits of posits versus floats for widely-used commercial applications.

Since the inception of posit data representation, there have been several implementations in the literature of posit arithmetic operations. The early and open-source hardware implementations of a posit adder and multiplier were presented in \cite{posit1} and \cite{posit2}. In \cite{posit2}, the authors covered the design of a parametric adder/subtractor while in \cite{posit1}, the authors presented parametric designs of float-to-posit and posit-to-float converters, and a multiplier along with the design of an adder/subtractor. A major disadvantage of the designs presented in \cite{posit1} and \cite{posit2} is that the designs are yet to be fully verified and contain multiple errors. 
The PACoGen open-source framework that can generate a pipelined adder/subtractor, multiplier, and divider is presented in \cite{posit3}. The design presented in \cite{posit3} has a disadvantage that of not synthesizing for the exponent size zero, and hence cannot be considered a fully parametric implementation. A more complete implementation of a parametric posit adder and multiplier generator is presented in \cite{iccd1}. 

\emph{Optical flow} is caused by the relative motion of an observer and a scene that has objects in motion. Out of several methods in the literature, we choose the Lucas-Kanade (LuKa) method for our experiments due to its simplicity and computational intensity \cite{of1}.

Recently, the open-source instruction set architecture (ISA) called RISC-V has gained a following in industry and academia. We integrated a posit adder and multiplier with the RI5CY core~\cite{pulp1} to create a posit-enabled RISC-V implementation. We compare area and energy numbers for field-programmable gate array (FPGA) synthesis of a RI5CY core with IEEE 754 compliant and with posit arithmetic.
The major contributions in this paper are as follows:

\begin{itemize}
    \item A detailed empirical study of LuKa using SoftPosit and SoftFloat where we compare numerical accuracy in LuKa for posits and IEEE 754 compliant floats
    \item RISC-V-based comparison of area and delay using posits versus fully-compliant IEEE 754 floats (to the best of our knowledge, this is the first such study)
    \item Performance analysis of LuKa on RISC-V with posit and IEEE 754 compliant floats, and discussion of current research issues in posit arithmetic
\end{itemize}
\vspace{-1mm}
The rest of the paper is organized as follows: In Section \ref{sec:back}, we present an overview of IEEE 754-2019 format, posit number format, and the LuKa method along with the relevant literature. In Section \ref{sec:analy}, accuracy analyses of LuKa using SoftFloat and SoftPosit are discussed in detail. A hardware implementation is presented in Section \ref{sec:imp} along with performance measurements. We summarize our conclusions in Section \ref{sec:con}.
\section{Background}\label{sec:back}
\subsection{IEEE 754 Compliant and Posit Number Systems}\label{sec:posit_sub}

\begin{figure}[!t]
\centering
\includegraphics[width = \columnwidth]{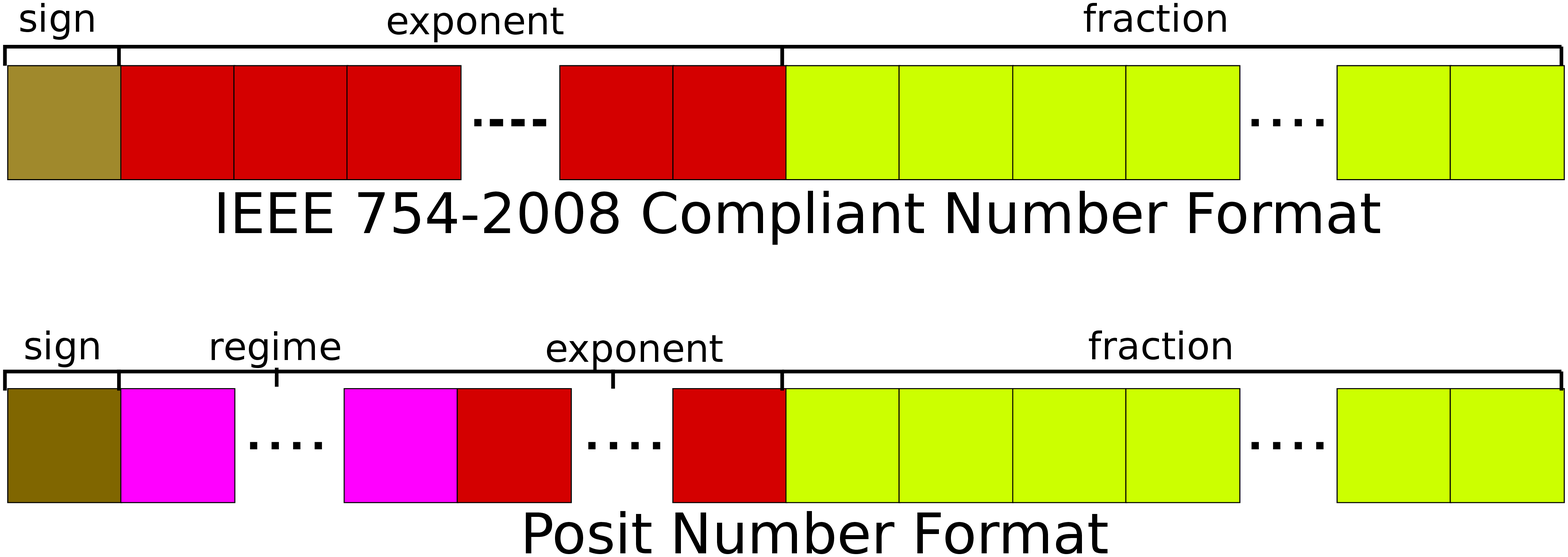}
\caption{Generic comparison of IEEE 754 floating-point (\emph{float}) and \emph{posit} number formats for non-exception values}
\label{fig:ieee754}
\end{figure}

The IEEE 754-2019 binary floating-point format numbers have three parts for \textit{normal} floats: a sign, an exponent, and a fraction (see Fig. \ref{fig:ieee754}). The sign is the most significant bit and indicates whether the number is positive or negative. In single precision, the next $8$ bits represent the exponent of the binary number ranging from $-126$ to $127$. The remaining $23$ bits represent the fractional part. The format is:
\begin{equation}
\textit{val} = ( -1 )^{\textit{sign}} \times 2^{\textit{exp}-\textit{bias}} \times ( 1.\textit{ fraction} ) 
\end{equation}

When the exponent bits express the minimum (all \texttt{0} bits) or maximum (all \texttt{1} bits), an exception value is indicated. It is currently common for vendors to claim IEEE 754 compliance in their hardware while actually complying only for the case of normal floats. Full IEEE 754 compliance for exception cases, deemed to be rare, is seldom supported in hardware; instead, traps to software or microcode are used. This approach degrades both performance and security; data-dependent timing creates a side-channel security hole.

\textit{Posit arithmetic} was proposed as a drop-in replacement for IEEE 754 arithmetic in 2017 \cite{gustafson1}. Posit arithmetic has several advantages over IEEE 754 arithmetic: higher accuracy for the most commonly-used values, simpler hardware implementation, smaller chip area, and lower energy cost~\cite{iccd1}~\cite{ihsen1}. Unlike IEEE 754 floats, there are no subnormal posit numbers, nor is there any need for them; $|x-y|$ produces a zero result if and only if $x=y$. There are only two exception cases: zero and not-a-real (NaR). For all other cases, the value \textit{val} of a posit is given by 
\begin{align}
    \textit{val}  = & (-1)^{\textit{sign}} \times \textit{useed}^{k} \times 2^\textit{exp}\times (1+\sum_{i=1}^{\textit{fn}-1}b_{\textit{fn} - 1 - i}2^{-i})
\end{align}
The regime indicates a scale factor of $\textit{useed}^{k}$ where $\textit{useed} = 2^{2^\textit{es}}$ and \textit{es} is the exponent size. The numerical value of $k$ is determined by the \emph{run length} of \texttt{0} or \texttt{1} bits in the string of regime bits. Run-length encoding of the regime automatically allows more fraction bits for the more common values for which magnitudes are closer to 1, and thus provides tapered accuracy in a bit-efficient way that preserves ordering. Further details about the posit number format and posit arithmetic can be found in \cite{gustafson1}. The posit format is depicted in Fig. \ref{fig:ieee754}.
\vspace{-3mm}
\subsection{Lucas-Kanade Method}
Despite its limitations in determining optical flow information in uniform images, the LuKa technique and its variants are the widely used methods for estimation of the optical flow in commercial products \cite{luka1}. Suppose $I$ is the brightness of the pixel at position $(x(t), y(t))$ at time $t$. We wish to solve
\begin{align}
    I_xu_x+I_yu_y+I_t = 0
\end{align}
where $I_x$, $I_y$, and $I_t$ represent the $x$, $y$, and $t$ directional gradients, respectively, and $u_x$ and $u_y$ represent the optical flow to calculate. To solve this equation, a \textit{local smoothness} constraint is added, which assumes the change in $u_x$ and $u_y$ in a small neighborhood of pixels to be extremely small. The final vector $\vec{u}$ containing the flow components is obtained from the equation
\begin{align}
    \vec{u} = (A^TA)^{-1} A^TB
\end{align}
where $A$ is the directional derivative matrix of the image and $B$ is the time derivative vector. The derivatives here are simple deltas from one image to the next with a resolution of 1/255. Since, the matrix $A{^T}A$ is a $2\times 2$ matrix, we use Cramer's rule for the matrix inversion.  
\subsection{Related Work}
There have been several attempts of posit implementation since the first proposal. The early parameterized designs were presented in~\cite{posit1},~\cite{posit2},~\cite{posit3}, and~\cite{iccd1}. The designs presented in~\cite{posit1},~\cite{posit2}, and~\cite{posit3} are open-source but do not synthesize for exponent size \emph{zero} while the design presented in~\cite{iccd1} is not open-source. A power-efficient posit multiplier is presented in~\cite{mul1}. The authors in~\cite{mul1} present a scheme where they divide the fraction part of the multiplier into several chunks and use them efficiently resulting in 16\% power efficiency over the base-line implementation. 

Several posit implementations are explicitly focused on machine learning applications. Performance-efficiency trade-off for deep neural network (DNN) inference is presented in~\cite{ml1}. The authors have discussed overall neural network efficiency and performance trade-offs in~\cite{ml1}. A template-based posit multiplier is presented in~\cite{ml2} where authors have incorporated training and inference of the neural networks. Authors have shown that 8-bit posits are as good as floats in inference. The \emph{Deep Positron} DNN architecture presented in~\cite{ml3} shows trade-offs between performance and hardware resources. The Deep Positron architecture uses an FPGA-based soft core to control the multiply-accumulate unit hardware (fixed-point, floating-point, and posit). The \emph{Cheetah} framework presented in~\cite{ml4} incorporates mixed-precision arithmetic alongside support for the conventional formats.      

RISC-V integration of posit arithmetic hardwares are presented in PERI~\cite{peri}, PERC~\cite{perc}, and Clarinet~\cite{clarinet}. PERI presented in~\cite{peri} uses SHAKTI C-class core as a base to attach posit arithmetic hardware, first as a tightly-coupled unit, and then as an accelerator connected through \emph{rocket custom coprocessor} (RoCC) interface. PERC presented in~\cite{perc} delves into a similar aspects while using \emph{RocketCore} as a base. Flute RISC-V core from Bluespec Inc is used for posit arithmetic experimentation in Clarinet~\cite{clarinet} where Melodica is the tightly-coupled posit core. Clarinet has a unique feature that it supports the \emph{quire} register as well for exact dot products; fused multiply-accumulation is a special case of the accumulation in the quire register. There also exists a couple of commercial attempts, such as the CRISP core by Calligo Technologies~\cite{calligo} and VividSparks~\cite{vivid}.

Very few implementations in the literature focus on application-specific posit arithmetic tuning wherein extensive analyses are performed before delving into the hardware designs. In our approach, we first emphasize application analyses followed by RISC-V integration of posit arithmetic unit. For our implementation of a posit adder and multiplier, we have used the improvised implementations of the designs proposed in~\cite{iccd1}, and for the divider, we have used the design presented in~\cite{posit3}. 

\section{LuKa using SoftPosit and SoftFloat}\label{sec:analy}
\begin{figure}[!t]
\centering
\includegraphics[width = \columnwidth]{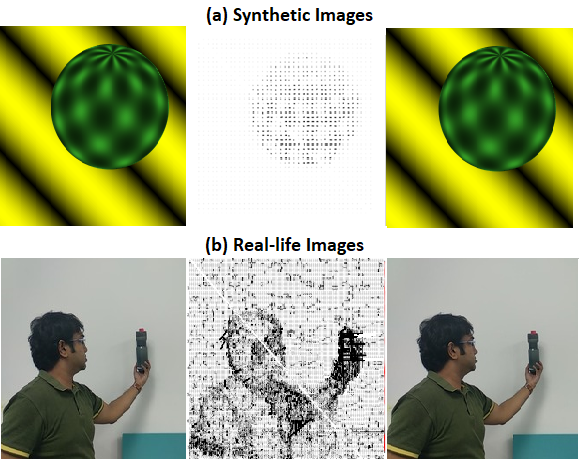}
\caption{Optical flow in consecutive frames in (a) synthetic images, and (b) real-life images. Images on the left and right side are the consecutive input images and the middle images represent the optical flow. None of the images are manipulated to support any particular number format.}
\label{fig:fig2}
\end{figure}

The study is conducted with synthetic images of a sphere (slightly rotated in each successive frame) and real-life images a human being (slightly translated in each successive frame), as shown in Fig.~\ref{fig:fig2}a and~\ref{fig:fig2}b. It is ensured that the images are well textured and the motion is very small for consecutive frames, eliminating the need to use regularization-based methods or multi-scale estimation. 

To perform the error analysis, LuKa is implemented in the C programming language. We ensure that the implementation has no dependency on any third-party or open source libraries. The \textit{reference} implementation uses double precision floating-point arithmetic. The code is executed with all consecutive pairs of frames as inputs over the whole data set of images and the optical flow values are obtained.


\begin{figure*}[!t]
\centering
\includegraphics[width = 1.8\columnwidth]{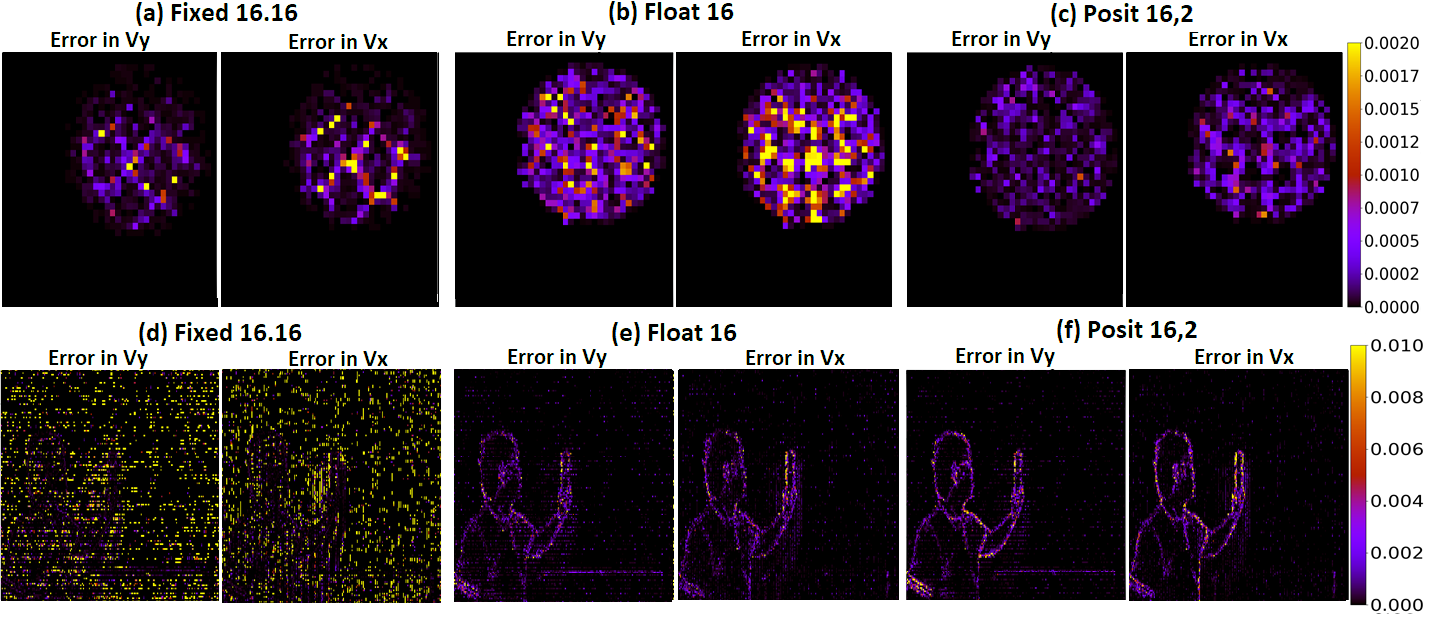}
\caption{Error heat maps for synthetic ((a), (b), and (c)) and real-life ((d), (e), and (f)) images in $y$ and $x$ for fixed-point, float, and posit formats (IEEE 754-2008 64-bit reference)}
\label{fig:fig3}
\end{figure*}

 \vspace{-3mm}
\subsection{Accuracy analysis for $16$-bit floats, fixed-point, and posits}
A primary goal of this study was to compare low-precision ($16$-bit) posit and float result accuracy. We also test a $16$-bit fixed-point format. The grey-scale pixel values ($0$ to $255$) are first scaled by dividing by \textit{norm}; we tested norm values ranging from $1$ to $255$. Each format has a preferred norm; for example, a too-small norm for floats leads to catastrophic overflow in the matrix multiplication step of the algorithm, since the largest real value they can represent is $65504$.

 For all three formats, we compare the results with a reference result. We pick the norm value that gives the smallest absolute error. Heat maps of the absolute error for both $u$ and $v$ are generated to visualize the distribution of error (Fig.~\ref{fig:fig3}). The heat maps and data presented are for the errors in the optical flow between two particular frames selected from the synthetic and real-life image data sets, representative for the whole experiment. 
 
For the $16$-bit (half-precision) float study, we use the Berkeley SoftFloat library by John Hauser, which provides an excellent stable software implementation of this precision that conforms to the IEEE 754 Standard. All optical flow values are calculated with $16$-bit floating-point variables and operations. For the $16$-bit floats, the best norm value is found to be $32$ for both synthetic and real-life images (We discuss the cause and implication of this in detail later in this section).

For the fixed-point implementation, we take advantage of the \textit{libfixmath} fixed-point math library. As with SoftFloat, a Q16.16 implementation of the code is prepared and executed on the same input data set. Heat maps (Fig.~\ref{fig:fig3}) are generated for the best cases (norm factors of 28 and 18 for the synthetic and real-life images respectively).

\begin{figure*}[!t]
     \centering
     \begin{subfigure}[b]{0.45\textwidth}
         \centering
         \includegraphics[width=\textwidth]{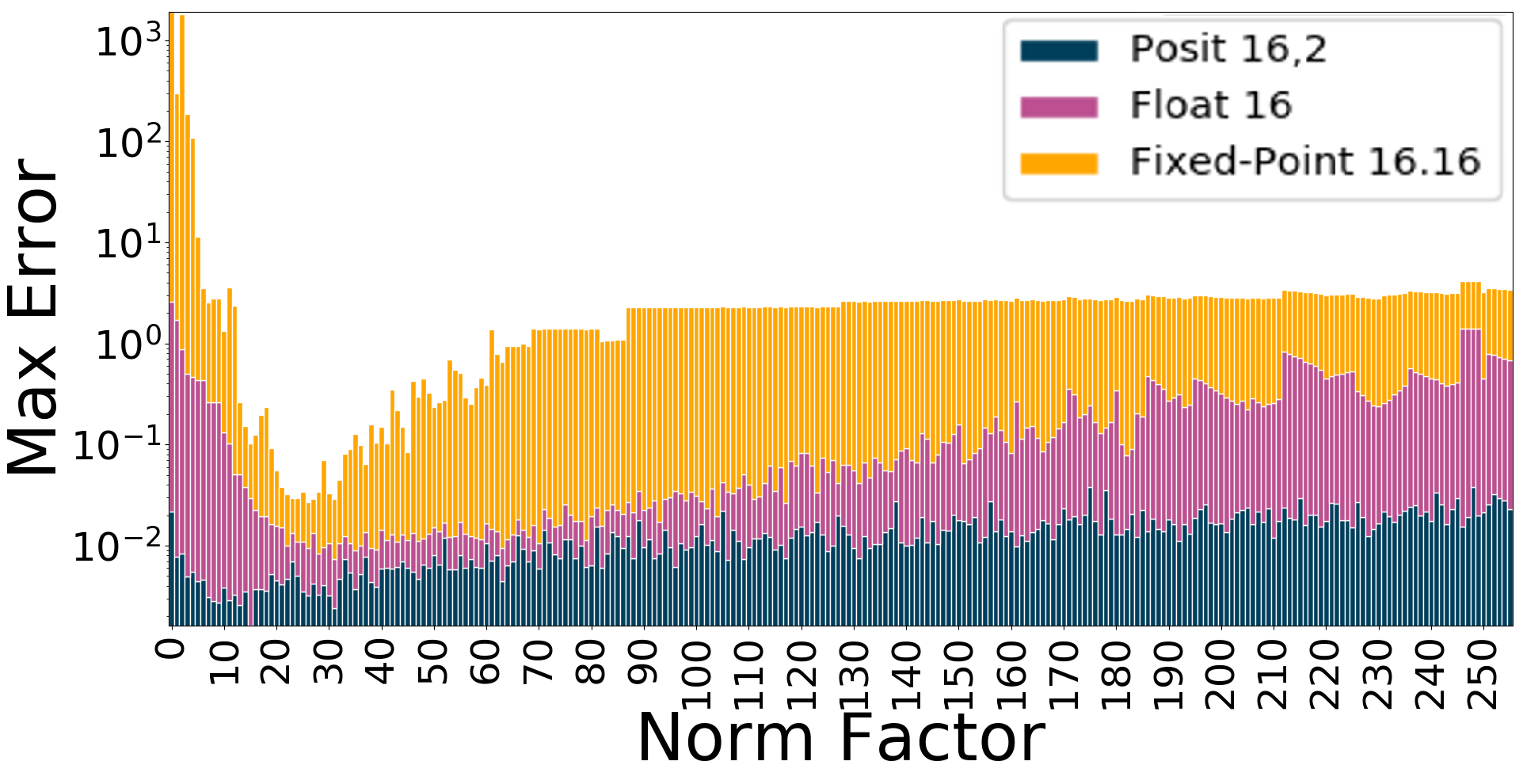}
         \caption{}
         \label{fig:norm}
     \end{subfigure}
     \begin{subfigure}[b]{0.45\textwidth}
         \centering
         \includegraphics[width=\textwidth]{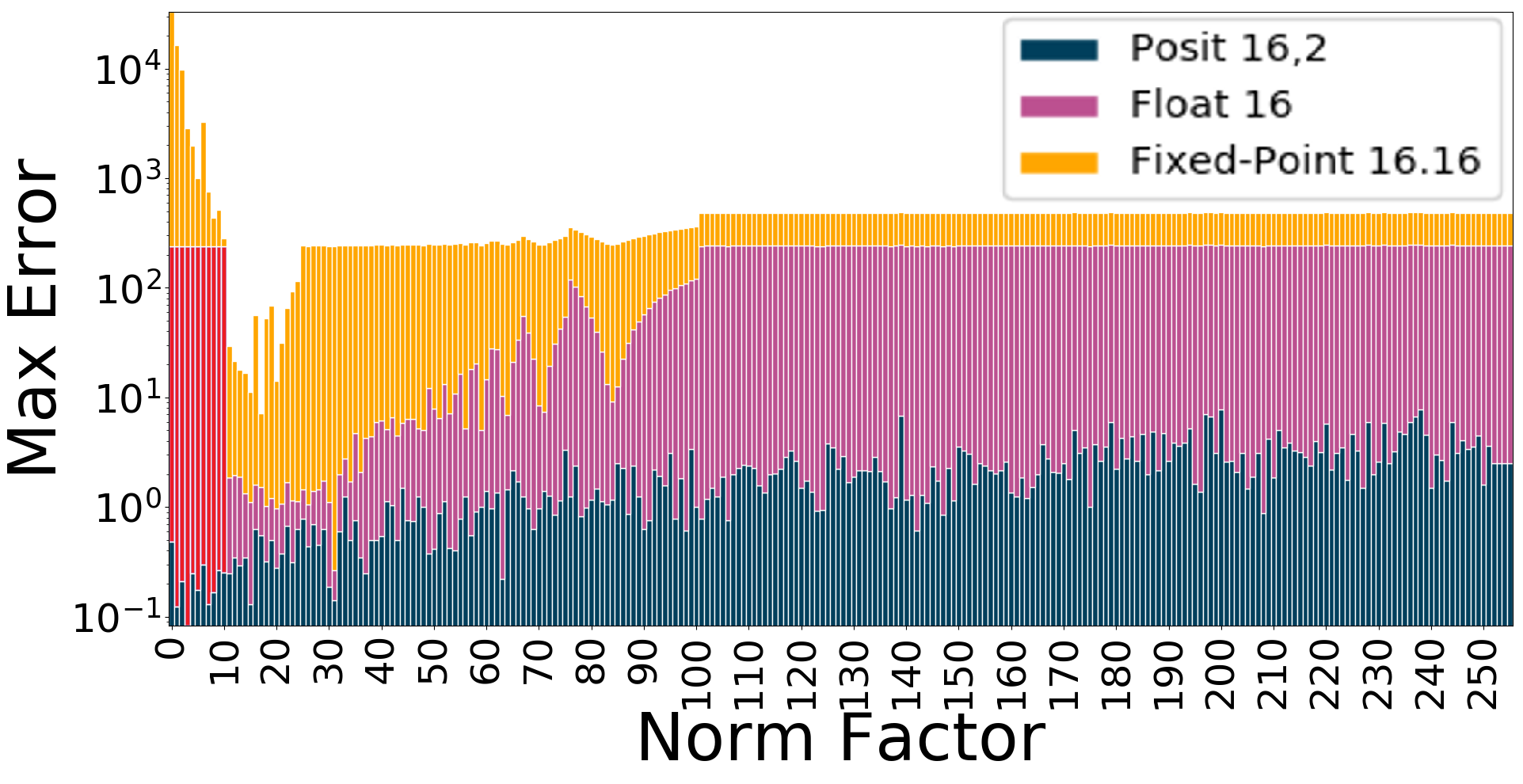}
         \caption{}
         \label{fig:norm_real}
     \end{subfigure}
        \caption{Trend in accuracy for different normalization factors in (a) synthetic images and (b) real-life images}
        \label{fig:fignormformats}
\end{figure*}

\begin{table}[!b]
 \scriptsize
    \caption{Absolute errors in optical flow}
    \vspace{-3mm}
	\begin{center}
		\begin{tabular}{ |p{2.6cm}|p{1.1cm}|p{0.85cm}|p{0.98cm}|p{1.0cm}|  }
			\hline
			&Fixed16.16 &Float 16 & Posit 16,1 & Posit 16,2 \\
			\hline
			Max Error (synthetic)&0.01579 &0.0047 &0.00272 &0.00163 \\
			\hline
			RMS Error (synthetic)&0.00057 &0.00049 &0.00016 & 0.00015 \\
			\hline
			Std. Deviation (synthetic)&0.00056 &0.00046 &0.00015 &0.00046 \\
			\hline
			Max Error (real-life) &5.6692 &0.125 &0.13412 &0.08333 \\
			\hline
			RMS Error (real-life) &0.12940 &0.00109 &0.00234 & 0.00108 \\
			\hline
			Std. Deviation (real-life) &0.12885 &0.00108 &0.00233 &0.00107 \\
			\hline
		\end{tabular}
		\label{lukasum}
	\end{center}
\end{table}

The posit implementation uses Cerlane Leong's SoftPosit library. It supports two $16$-bit configurations with $\textit{es} = 1$ and $\textit{es} = 2$; we found $\textit{es} = 2$ the better fit for this application. The code was ported with all variables, and operations changed to posits. The use of the \emph{quire}, supported by the SoftPosit library, is out of the scope of this work, but in future work may further improve the accuracy in the matrix multiplication step. Pixel values are again normalized. Optical flow values and errors are calculated as before. Norm values of $16$ and $4$ present the smallest error for synthetic and real-life images respectively. 

Table \ref{lukasum} summarizes the results obtained. ``posit 16,$k$'' refers to 16-bit posits with $\textit{es}=k$. For the synthetic images, the maximum error for the fixed-point format is an order of magnitude higher than the other formats. However, this is not the case with RMSE which is very close to the RMSE for float 16, albeit $\sim4\times$ more than posit 16,2. This is also evident from visualizing the heat maps and confirms that fixed-point format gives mostly accurate values with few of very high absolute error. Posit 16,2 has $\sim3\times$ lower maximum and RMS errors compared to float 16 while posit 16,1 has an error profile intermediate to the two formats. 

For the real-life images, results are slightly different. It should be noted that no additional filters were applied to the images before the optical flow calculations and they were extracted from video as-is. They lack the texture and sharpness of synthetic images and are noisier in general. It is found that both the maximum and RMS errors for fixed-point format in this case are \textit{two} orders of magnitude higher compared to floats and posits. Float 16 performs equivalent to posit 16,2 and better than posit 16,1 in terms of RMS error, although, the max error for float is $\sim1.5\times$ the max error for posit 16,2.   

The summary in Table \ref{lukasum} presents the best-in-class results, but for a more generic view of the performance of the formats, the max errors are plotted against the normalization factors in Fig.~\ref{fig:fignormformats} in the form of bar charts. Normalizing by $16$ gives very high accuracy for posits in both data sets (best for synthetic and next-best for real-life). In other words, scaling the original pixel values from ($0$--$255$) to ($0$--$16$) leads to further improvement in result accuracy. This is because of the \emph{tapered accuracy} property of posits; accuracy is maximized for values close to $0$ in magnitude. Dividing by $16$ centers the (nonzero) pixel values $x$ in the range $\tfrac{1}{16}\leqslant x<16$. Posit 16,2 has its maximum accuracy in exactly this range, $12$ significant bits. Float 16 is consistently less accurate than posit 16,2 and fixed-point is consistently less accurate than both floats and posits. The red bars in Fig.~\ref{fig:fignormformats} (b) indicate NaN float values that are generated for norm values in the range of 1 to 8 that are too small to prevent overflow. (Posit 16,2 can represent real values up to about $7.2\times10^{16}$.)

\begin{figure}[!t]
\centering
\includegraphics[width = 0.9\columnwidth]{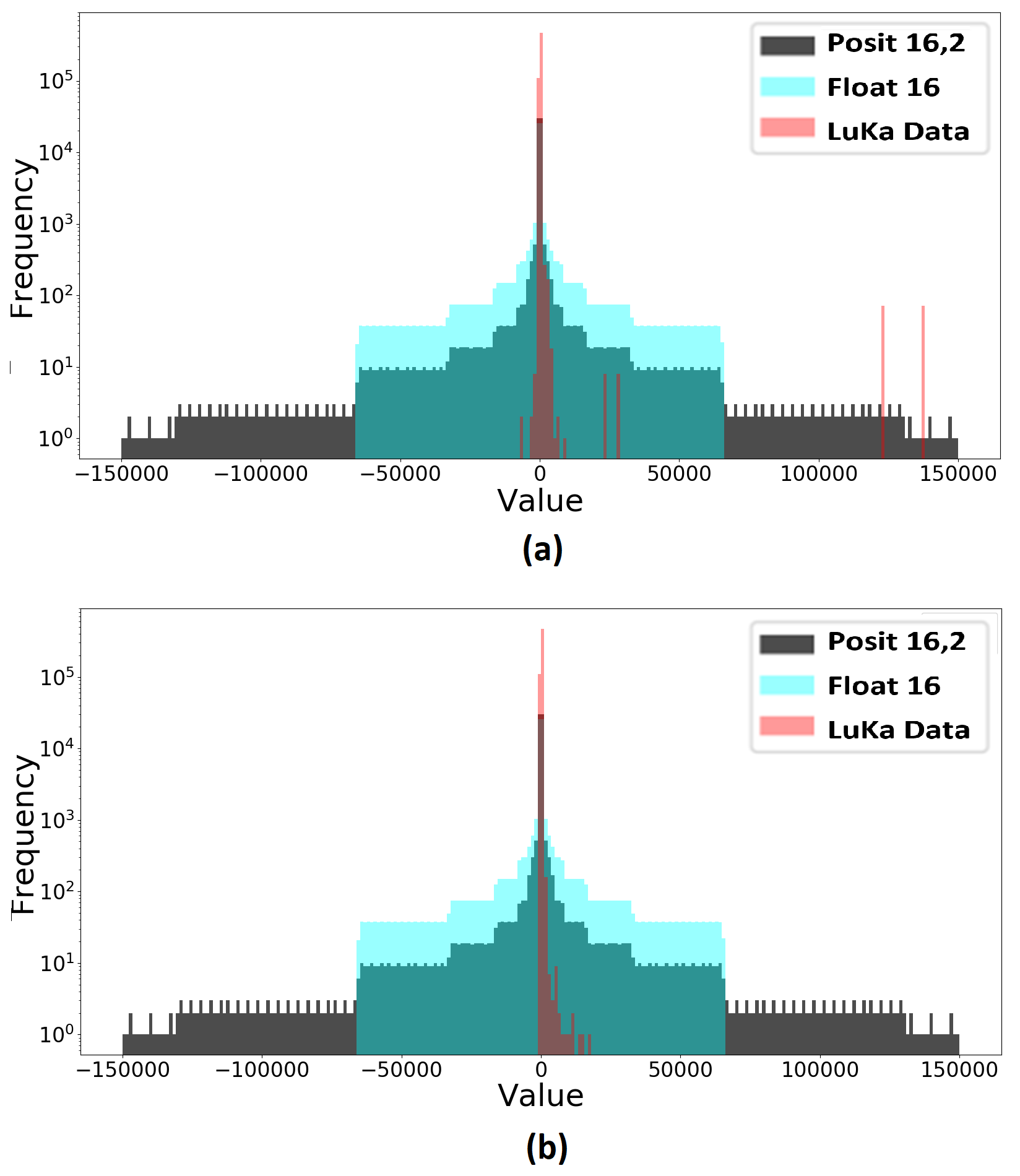}
\caption{Histogram overlap of posits, floats, and unique data values generated during reference (double precision) implementation run of LuKa with normalization factors of (a) 255 and (b) 32 for synthetic images} 
\label{fig:hist_syn}
\end{figure}

\begin{figure}[!t]
\centering
\includegraphics[width = 0.9\columnwidth]{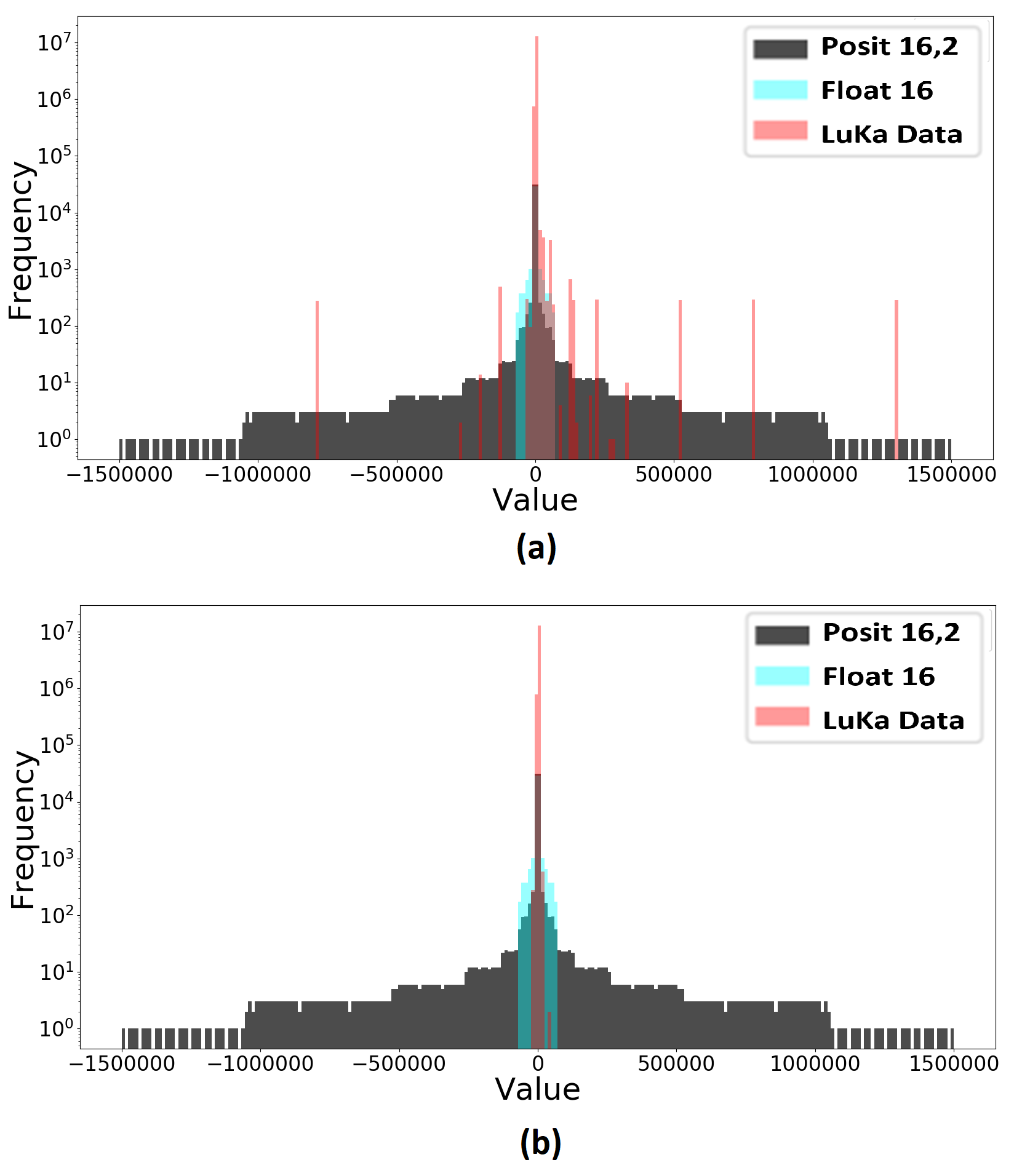}
\caption{Histogram overlap of posits, floats, and unique data values generated during reference (double precision) implementation run of LuKa with normalization factors of (a) 255 and (b) 32 for real-life images} 
\label{fig:hist_real}
\end{figure}

Next, we delve deeper into the float 16 and posit 16,2 formats to understand why float 16 performs so well in certain regions (such as $\text{norm}=32$ for floats). Data values generated from each and every intermediate arithmetic operation performed in the LuKa algorithm is collected in the reference implementation for norm values of $255$ (scaling pixels to range from $0$ to $1$) and $32$ (scaling pixels to range from $0$ to $8$). This is done for both synthetic and real-life images. From this intermediate data, all the unique values are extracted and analyzed. It is found that normalizing by $32$ limits the dynamic range of the data values generated during the calculation, bringing them within the dynamic range of float 16 (Fig.~\ref{fig:hist_syn} and Fig.~\ref{fig:hist_real}). Fig.~\ref{fig:hist_syn} and ~\ref{fig:hist_real} also present overlapped histograms of float 16, posit 16,2 and the unique data values generated. A good overlap entails a better number system for the application in hand. Posits have a far wider dynamic range than floats and hence perform better in general across all norm factors. For the norm factor of $32$ where float 16 has adequate dynamic range, the tapered nature of data (with high density of values around $0$) gives a slight edge to posits resulting in a marginally lower error at that norm, though not as low as posits using their optimum norm. Fig.~\ref{fig:hist_syn} also shows that a relatively smaller error in the larger data values carries more weight in the final result accuracy than the larger error in smaller data values. However, a deeper study with more applications is needed to substantiate this claim. This study shows the advantages of using posits over other formats for calculating the optical flow using the LuKa method.        

\section{Hardware implementation of LuKa}\label{sec:imp}
\begin{figure}
  \centering
  \includegraphics[width=\columnwidth]{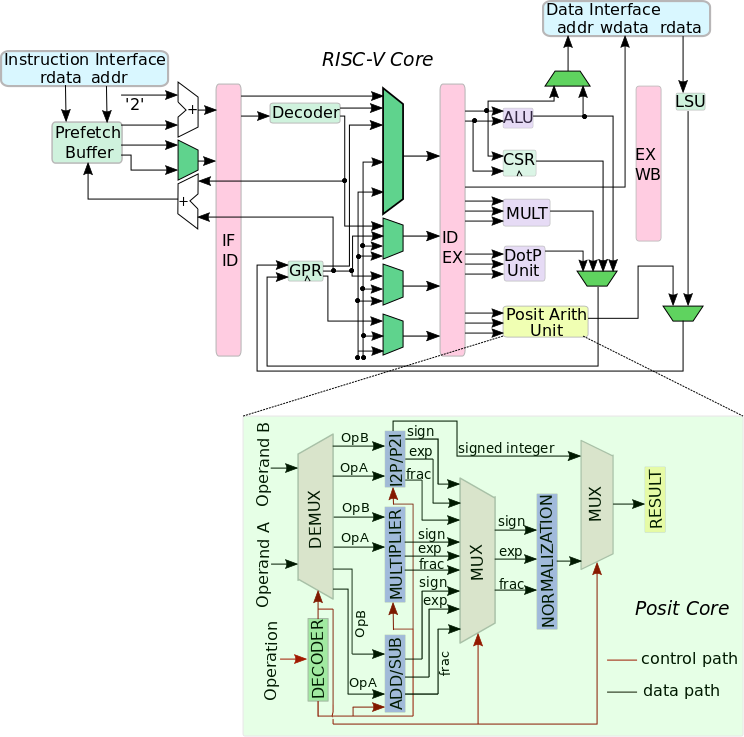}  
  \caption{RISC-V integration of posit core}
  \label{fig:fig6}
\end{figure}

\begin{table}[!b]
\scriptsize
\caption{Adder, multiplier synthesis results (delays in ns)}
    \vspace{-2mm}
	\begin{center}
\begin{tabular}{|p{0.6cm}|p{0.85cm}|p{0.7cm}|p{0.7cm}|p{0.85cm}|p{0.7cm}|p{0.7cm}|}
	\hline
	\centering
    & \multicolumn{3}{|c|}{Adder} & \multicolumn{3}{|c|}{Multiplier} \\ \cline{2-7}
	($n$,\textit{es}) & LUT & \makecell{Logic\\Delay} & \makecell{Net\\Delay}  & LUT & \makecell{Logic\\Delay} & \makecell{Net\\Delay} \\
	\hline
	(8,0)	&185 (0)	&8.83	&21.12 &95 (1)	&7.43	&13.13\\
	\hline
	(8,2)	&181 (0)	&9.68	&20.92 &96 (0)	&4.28	&14.07\\
	\hline
	(16,1) &400 (0)	&12.77	&19.01  &229 (1)	&10.16	&13.55\\
	\hline
	(16,2)	&391 (0)	&14.78	&20.07 &226 (1)	&10.76	&13.09\\
	\hline
	(32,2)	&866 (0)	&17.30	&24.57 &572 (4)	&15.55	&16.38\\
	\hline
\end{tabular}
\label{tab:table1}
\end{center}
\end{table}

\begin{table}[!b]
\scriptsize
	\caption{FPGA synthesis results for FPU and Posit cores}
	\vspace{-2mm}
	\begin{center}
		\begin{tabular}{ |p{1.5cm}|p{1.5cm}|p{2.3cm}|  }
			\hline
			& LUT Count & Delay (ns)\\
			\hline
			FPU~\cite{pulp1}& 2669 &50 (20 MHz)\\
			\hline
			Posit (16,1)& 2082 &55 (18.18 MHz) \\
			\hline
			Posit (16,2)& 2024 &42 (23.81 MHz) \\
		    \hline
			Posit (28,2)& 2780 &71 (14.08 MHz) \\
			\hline
			Posit (32,2)& 2810 &71 (14.08 MHz) \\
			\hline
		\end{tabular}
		\label{tab:riscy2}
	\end{center}
\end{table}


\begin{table}[!b]
\scriptsize
\caption{LUT count comparison}
    \vspace{-2mm}
	\begin{center}
\begin{tabular}{|p{0.45cm}|p{0.30cm}|p{0.27cm}|p{0.38cm}|p{0.3cm}|p{0.70cm}|p{0.80cm}|p{0.80cm}|p{0.74cm}|}
	\hline
    & \multicolumn{4}{|c|}{Adder} & \multicolumn{4}{|c|}{Multiplier} \\ \cline{2-9}
	($n$,\textit{es}) & Ours & \cite{iccd1} & \cite{posit3} & \cite{florent}  & Ours & \cite{iccd1} & \cite{posit3} & \cite{florent} \\
	\hline
	(8,0)& 185 & NS &NS &NR& 95\hspace{0.3mm}(1)& NS&NS &NR\\
	\hline
	(8,2)& 181 & 208 & 196& NR& 96\hspace{0.3mm}(0)& 131\hspace{0.3mm}(0) & 123\hspace{0.3mm}(0)&NR\\
	\hline
	(16,1)& 400 &391 &460&320& 229\hspace{0.3mm}(1)& 218\hspace{0.3mm}(1) &271\hspace{0.3mm}(1)&253\hspace{0.5mm}(1)\\
	\hline
	(16,2)& 391 & 404 &492&NR& 226\hspace{0.3mm}(1)& 223\hspace{0.3mm}(1) & 272\hspace{0.3mm}(1)&NR\\
	\hline
	(32,2)& 866 & 981 &1115&745&572\hspace{0.3mm}(4)& 572\hspace{0.3mm}(4) &648\hspace{0.3mm}(4)&469\hspace{0.5mm}(4)\\
	\hline
\end{tabular}
\label{tab:table2}
\end{center}

 \end{table}

We integrate a verified posit adder and multiplier to the RI5CY core of the \textit{Pulpino} platform~\cite{pulp1}. We disintegrate the existing floating-point unit (FPU) from the RI5CY core and integrate the posit arithmetic unit (PAU) generated adder and multiplier in the core, as shown in Fig. \ref{fig:fig6}. RI5CY is a 32-bit core based on RISC-V ISA supporting floating-point instructions. We generate a 32-bit adder and multiplier using PAU for the integration.
The developed parametric posit hardware generator allows us to choose any posit configuration ($n$ or \textit{es}) and generate \textit{adder}, \textit{multiplier}, \textit{integer to posit converter} (int2pos) and \textit{posit to integer converter} (pos2int) hardware operators. The PAU has been exhaustively tested against the SoftPosit library for $(n, \textit{es})=(8,0)$, $(7,2)$, $(8,2)$, $(9,2)$, $(10,2)$, $(11,2)$, and $(12,2)$ configurations. Furthermore, we have also tested for $(16,1)$ configuration for $\sim1.31$ billion combinations ($\approx$ $31\%$), mainly covering the corner cases.

Based on the conclusions obtained from the optical flow study, we synthesize and integrate a $(16,2)$ PAU core with Pulpino. The results obtained post-integration are shown in Table \ref{tab:riscy2}. The baseline version of the RI5CY core is the version with the native IEEE 754 FPU. Switching to the $(16,2)$ configured PAU core affords enormous savings in data RAM usage at a tolerable loss in accuracy for our optical flow application. Integration results for other configurations of the posit core are also provided for reference. The table also shows the FPGA  delay for various configurations of the PAU core. $16$-bit versions show a delay of $55$~ns and  $42$~ns, comparable to the $50$~ns achieved by the baseline FPU from Pulpino. 

Table \ref{tab:table1} presents the detailed synthesis results of the PAU adder and multiplier. This PAU is synthesized for Zedboard with Xilinx Zynq-7000 SoC. Vivado 2018.3 is used for the FPGA synthesis results. Both the adder and multiplier implementations are purely combinational in nature and without any pipelining. The DSP counts are given in parentheses next to LUT counts for all the configurations. In general, reasonably good area and delay numbers are observed. We benchmark our PAU against the other published results on posit hardware in Table \ref{tab:table2}. NS marks configurations that are not synthesizable, and NR signifies not reported in the paper. Again, the DSP counts are given in parentheses next to the LUT counts for multiplier. The adders do not use any DSP blocks across all the implementations. The LUT count of this work shows a significant improvement over existing parametric posit hardware generators \cite{posit3, iccd1}. It is also more extensively tested compared to these previous works. \cite{florent} shows lower area footprint and is a good candidate for our future implementations. To the best of our knowledge,~\cite{florent} and this work are the best published implementations of a \textit{parametric} posit hardware generator.



\section{Conclusion}\label{sec:con}
The purpose of this work is to analyze the benefits and shortcomings of posit arithmetic over floats in real-world applications. We have demonstrated the clear advantage of using posits instead of floats for calculating optical flow using LuKa. An order of magnitude improvement in accuracy is observed when the algorithm is implemented using posits instead of floats in synthetic images. In contrast, for real-life images, the accuracy is comparable. A fixed-point approach has accuracy too low to be viable. The algorithm was then further implemented in hardware on a RISC-V core that has been modified to support posit 16,1 and 32,2. The synthesis results of the modified core, as well as the Posit Arithmetic Unit, were presented; Pulpino performs well with a lower LUT count than the single-precision FPU. Our PAU is also shown to be comparable (if not better) in terms of area, to other state-of-the-art posit hardware.

\bibliographystyle{IEEEbib}
\bibliography{IEEEabrv,refs}

\end{document}